\documentclass[preprint,12pt]{elsarticle}

\usepackage[british]{babel}
\usepackage{hyphenat}
\usepackage{lineno,hyperref}
\modulolinenumbers[5]

\usepackage{graphicx,amssymb}
\usepackage{xcolor}
\usepackage{float}
\usepackage{booktabs}
\usepackage{tabularx}

\usepackage{color}

\journal{Materials Chemistry and Physics}

\begin{document}

\begin{frontmatter}



\title{Microwave-assisted synthesis and characterization of undoped and manganese doped zinc sulfide nanoparticles}


\affiliation[CFI]{organization={Institute of Solid State Physics, University of Latvia},
            addressline={Kengaraga Street 8}, 
            city={Riga},
            postcode={LV-1063}, 
            country={Latvia}}

\author[CFI]{Alexei Kuzmin\corref{cor1}}
\cortext[cor1]{Corresponding author}
\ead{a.kuzmin@cfi.lu.lv}

\author[CFI]{Milena Dile}

\author[CFI]{Katrina Laganovska}

\author[CFI]{Aleksejs  Zolotarjovs}

\begin{abstract}
Undoped and Mn-doped ZnS nanocrystals were produced by the microwave-assisted solvothermal method and characterized by X-ray diffraction, photoluminescence spectroscopy and scanning electron microscopy with energy-dispersive X-ray spectroscopy. 
All samples have the cubic zinc blende structure with the lattice parameter in the range of $a$=5.406--5.411~\AA, and the average size of crystallites  is in the range of 6--9~nm. These nanoparticles  agglomerate and form large grains with an average size of up to 180~nm. 
The photoluminescence of the undoped ZnS sample shows a broad emission band located at 530~nm, attributed to the defects at the surface of nanoparticles.  In all Mn-doped samples, the emission peak at 598~nm was observed 
assigned to the characteristic forbidden transition between excited ($^4$T$_1$) and ground ($^6$A$_1$) levels of Mn$^{2+}$.
Synchrotron radiation X-ray absorption spectroscopy at the Zn and Mn K-edges combined with reverse Monte Carlo (RMC) simulations  based on the evolutionary algorithm confirms that manganese ions substitute zinc ions. However, the difference in the ion sizes ($R$(Mn$^{2+}$(IV)) = 0.66~\AA\ and $R$(Zn$^{2+}$(IV)) = 0.60~\AA) is responsible for the larger interatomic distances Mn--S (2.40(2)~\AA) compared to Zn--S (2.33(2)~\AA). The static structural relaxations in ZnS:Mn nanoparticles are responsible for the large values of the mean-square displacements factors  for Zn, S and Mn atoms obtained by RMC simulations.
\end{abstract}

\begin{keyword}
 ZnS:Mn  \sep X-ray diffraction \sep Scanning electron microscopy \sep Photoluminescence spectroscopy \sep X-ray absorption spectroscopy 
\end{keyword}

\end{frontmatter}


\section{Introduction}
\label{sec:sample1}

Zinc-based nanostructures have been of interest due to their low toxicity, eco-friendly nature, biocompatibility as well as and the possibility of using them in various forms, such as nanoparticles, nanowires, nanotubes, nano-wells, etc. \cite{Barman2020}. Among them, zinc sulfide (ZnS) is an important  binary II-VI semiconductor compound  with a direct wide band gap energy of 3.6-3.7~eV at room temperature \cite{Fan1983,Ves1990,Kurnia2015,Lin2017} and a high refractive index, as well as a high transmittance in the visible region \cite{Bhushan2019}.
ZnS has been widely used  in numerous optical applications such as ultraviolet light-emitting diodes \cite{Lee2008}, flat panel displays \cite{Liu2001} and thin-film electroluminescent devices \cite{Dimitrova2000,Mastio2000}. 

The doping of nanocrystals with optically active luminescence centers may create new opportunities in the study and application of nanoscale materials.
Zinc sulfide nanoparticles doped with  Mn$^{2+}$ isoelectronic impurities  have  been  the  most  efficient  electroluminescent  (EL)  phosphor  material  over  the years  and  have been  extensively  studied  for  their  enhanced  quantum  efficiency,  increased luminescence intensity and shortened life time \cite{Bhargava1994,Bol1998}. Moreover, they represent a great interest in  chemo/biosensing and bioimaging applications \cite{Wu2013}.

Various synthetic routes have been developed to prepare Mn-doped ZnS such as the chemical precipitation method \cite{Chandrakar2015}, organometallic  method \cite{Karar2004}, reverse micelle method \cite{Kubo2002}, and sol-gel method \cite{Biggs2009} as well as solvothermal \cite{Dong2012} for use in the aforementioned applications.  

However, in addition to the great quantum efficiency and luminescence intensity, Mn-doped ZnS is also a promising material for transparent thin-film electroluminescent displays \cite{Wood2009}. Although some commercial examples are available, large-scale manufacturing is still a significant challenge. To make the material more commercially viable, the method of synthesis needs to not only ensure the necessary characteristics but also be easily scalable and adaptable in industrial-scale thin-film manufacturing. One method that offers low cost, high quality, low agglomeration with small enough grain sizes (below 10~nm) is microwave-assisted solvothermal synthesis.

The microwave-assisted solvothermal method is known as a promising, simple, rapid and environmentally friendly method for nanomaterial synthesis. This method involves the main advantages of solvothermal synthesis such as reaching high pressures and temperatures in a reaction vessel in a relatively short time and preparing highly crystalline products with narrow particle size distribution and low agglomeration \cite{Nunes2019book}.

However, efficient mixing and convection in a reaction vessel are limited during regular solvothermal synthesis as a result of a high thermal gradient, so the mixture temperature rises close to the surface of the vessel where the vast majority of reactions take place \cite{Scajev2021}.

Therefore, microwave radiation is used to control interactions between materials. Under the electromagnetic field, the particles of materials can produce different types of polarization (electron, atom, orientation and space charge). This results in changes in some solvent properties, for instance, vapour pressure, density, viscosity and surface tension that allows one to transfer the microwave radiation energy evenly throughout the mixture and promote uniform reaction mixture heating in a short time with minimal heat losses \cite{Meng2016}.

In this study,  we focused on the properties of ZnS:Mn nanocrystals produced by the microwave-assisted solvothermal method and characterized by X-ray diffraction (XRD), photoluminescence spectroscopy and scanning electron microscopy (SEM) with energy-dispersive X-ray spectroscopy. Synchrotron radiation X-ray absorption spectroscopy (XAS) combined with reverse Monte Carlo (RMC) simulation was used to study the local atomic structure of nanocrystals with a particular emphasis on
relaxation phenomena around Mn-dopants.

\section{Materials and methods}

\subsection{Materials}

Zinc chloride (ZnCl$_2$, purity 99.9\%; Supelco) and sodium sulphide  (NaS$_2$$\cdot$9H$_2$O, purity 98\%; Acros Organics) were used as the main precursors for ZnS synthesis. Manganese acetate tetrahydrate (Mn(CH$_3$COO)$_2$$\cdot$4H$_2$O, purity 99.9\%; Sigma Aldrich) was used as a dopant source. The reaction was carried out in ethylene glycol (C$_2$H$_6$O$_2$, 99.5\%, Fisher Scientific) media. Methanol (CH$_3$OH, purity 99.9\%; Fisher Scientific) was used for the removal of synthesis reaction residue. All the chemicals were used without any further purification.

\subsection{Synthesis}

Undoped and Mn-doped (0.5, 1.0, 1.5, 2.0 at\%) powders were synthesized by microwave-assisted solvothermal (MWST) method. 
First, two solutions (A and B) were produced. Solution A was prepared by dissolving ZnCl$_2$ in 45~mL ethylene glycol. After its complete dissolution, Mn(CH$_3$COO)$_2$$\cdot$4H$_2$O was added in order to achieve a solution with 0.4~M total metal ion concentration and desired Zn and Mn ratio (0.5, 1.0, 1.5, and 2.0 at\%). Solution B was prepared by dissolving 0.018~mol NaS$_2$$\cdot$9H$_2$O in 45~mL ethylene glycol. Both solutions were stirred for 30~min. Then, 40~mL of solution A was slowly poured into the reaction vessel, after which 40~mL of solution B was slowly added to the mixture, and the reaction vessel was tightly sealed. Reaction mixtures for undoped ZnS powders were prepared in the same way without adding Mn precursor. 

The MWST synthesis was performed in a Milestone synthWAVE T660 (Milestone Srl) microwave reactor in an inert atmosphere (N$_2$, purity 99.999\%). During the reaction, the mixture was heated up to 200~$^\circ$C for 15~minutes, and further, the temperature remained constant for 1~h 25~min. Subsequently, the reaction mixture was cooled down for 30~min. The reactor was operated at 2.45~GHz frequency with power ranging from 0 to 100\% of full power (1.5~kW) and starting pressure 40~bar.

After synthesis, the resulting mixtures were centrifugated for 30~min, washed 8 times with 15~mL methanol,  and the precipitate was separated. Thus obtained samples were dried at 60~$^\circ$C temperature for 72~h and ground for further characterization.

\subsection{Characterization}

\begin{figure}[t]
	\centering
	\includegraphics[width=0.95\linewidth]{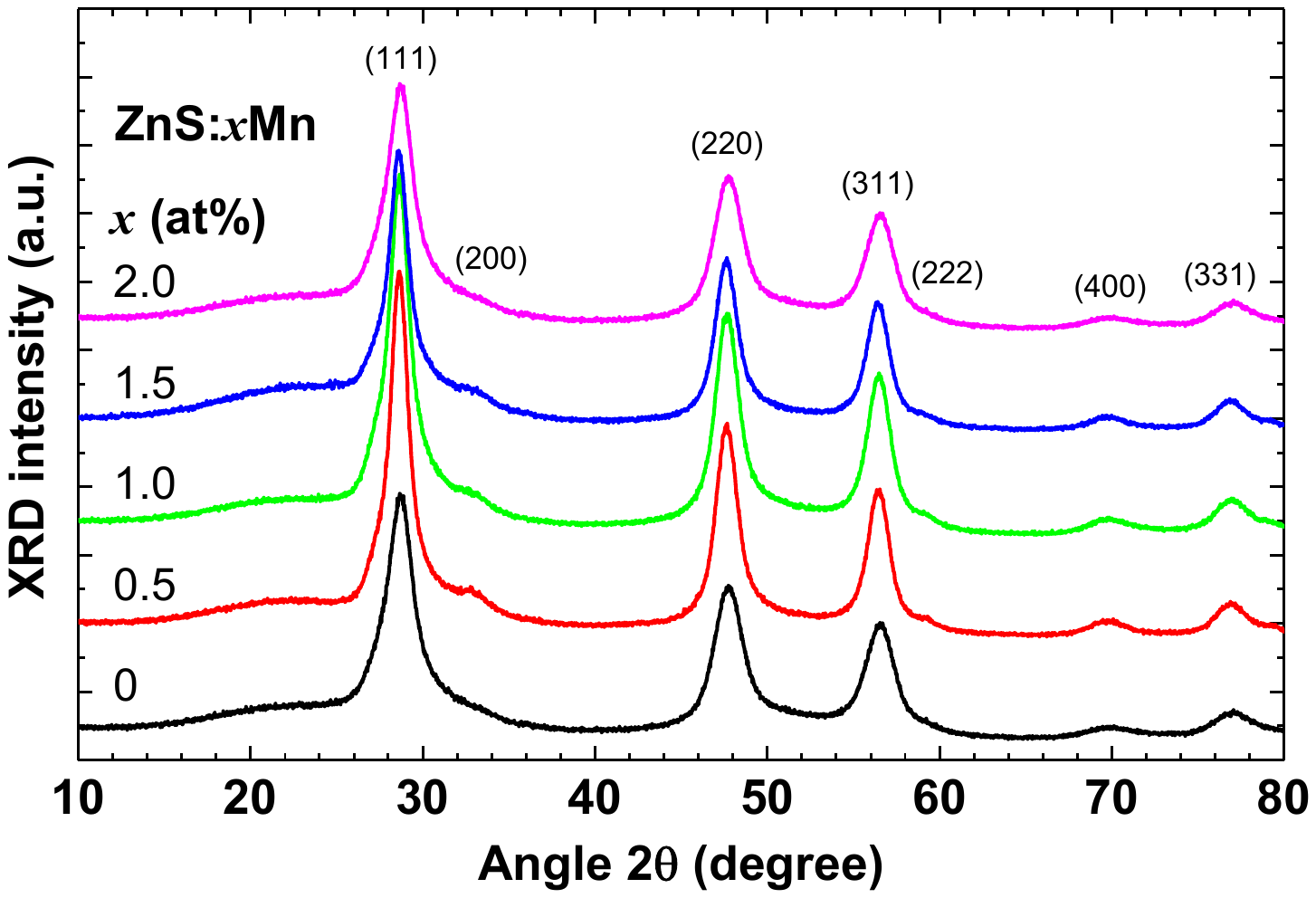}
	\caption{X-ray diffraction patterns of pure and doped ZnS:$x$Mn nanopowders ($x$ = 0, 0.5, 1.0, 1.5, 2.0~at\%), synthesized by the microwave-assisted solvothermal method. The crystalline planes of the cubic zinc blende structure are indexed.   }
	\label{fig1}
\end{figure}

The structure of pure and Mn-doped ZnS nanopowders was determined by the X-ray powder diffraction (XRD) technique. The XRD patterns (Fig.\ \ref{fig1}) were measured using a benchtop Rigaku MiniFlex 600 diffractometer with Bragg-Brentano $\theta$-2$\theta$ geometry equipped with the 600 W Cu anode (Cu K$\alpha$ radiation, $\lambda$ = 1.5406~\AA) X-ray tube operated at 40~kV and 15~mA. Rietveld refinement (Table\ \ref{table1}) was performed with the BGMN \cite{Taut1998} software using the Profex code \cite{Doebelin2015}.

\begin{table}[t]\footnotesize
	\caption{Rietveld analysis of the XRD patterns of ZnS:$x$Mn nanopowders. The Mn content $x_{\rm EDX}$ according to the EDX analysis is also given. } \label{table1}
	\renewcommand{\arraystretch}{1.2} 
	\begin{tabular}{@{}llcc}
		\hline
		$x$ (at\%) & $x_{\rm EDX}$  (at\%) 	& Lattice parameter $a$ (\AA)	& Crystallite size $d$ (nm)   \\
		\hline
		0	&0    &5.406(2)	&5.8(2) \\
		0.5	&0.6  &5.410(2)	&8.8(2) \\
		1.0	&1.3  &5.408(2)	&6.9(2) \\
		1.5	&1.4  &5.411(2)	&7.5(2) \\
		2.0	&1.0  &5.406(2)	&5.8(2) \\
		\hline
	\end{tabular}\\[2pt]
\end{table}

Scanning electron microscope (SEM) Thermo Fisher Scientific Helios 5 UX was used to examine morphologies and microstructure  of all samples in 
secondary electron (SE) mode. The elemental composition was studied by energy-dispersive X-ray spectroscopy (EDX).  SEM was operated for imaging at 2~kV voltage and during the elemental analysis at 30~kV. 

Photoluminescence measurements were performed using a YAG laser FQSS266 (CryLas GmbH) 4th harmonic at 266~nm (4.66~eV) at room temperature. All samples were pressed into tablets of equal size, which made it possible to compare the intensity between the samples. Luminescence spectra were recorded using
an Andor Shamrock B-303i spectrograph equipped with a CCD camera (Andor DU-401A-BV).

X-ray absorption spectroscopy of pure and Mn-doped (1 and 2~at\%) ZnS nanopowders was performed with the aim to determine the local environment of Mn ions and their influence on ZnS structure. X-ray absorption experiments were performed in transmission and fluorescence modes at the P65 Applied XAFS beamline \cite{Welter2019} of the HASYLAB/DESY PETRA III storage ring. A fixed-exit double-crystal Si(111) monochromator was employed at the Zn and Mn  K-edges, and the harmonic reduction was achieved using the uncoated silicon plane mirror. X-ray absorption spectra were collected in transmission mode at the Zn K-edge using two ionization chambers, whereas passivated implanted planar silicon (PIPS) detector was used in fluorescence mode at the Mn K-edge. To reduce thermal disorder, all measurements were performed at $T$=10~K using the SuperTran-VP (Janis Research Company, LLC) continuous flow liquid helium cryostat.

\subsection{EXAFS data analysis}

Experimental extended X-ray absorption fine structure (EXAFS) spectra of pure and Mn-doped (1 and 2 at\%) ZnS nanopowders at the Zn and Mn K-edges were extracted using the XAESA code \cite{XAESA} following a conventional procedure \cite{Kuzmin2014}. Note that the peaks in the Fourier transforms (FTs) of EXAFS spectra are located at distances that are slightly shorter than their crystallographic values because the FTs were not corrected for the phase shift present in the EXAFS equation.

The analysis of EXAFS spectra was performed by the reverse Monte Carlo method as implemented in the EvAX code \cite{Timoshenko2012, Timoshenko2014}. The method is based on the evolutionary algorithm (EA), which accelerates the optimization procedure \cite{Timoshenko2014}, and uses the Morlet wavelet transform (WT) for a comparison of the experimental and calculated EXAFS spectra simultaneously in $k$ and $R$ spaces \cite{Timoshenko2009}. Such an approach was successfully used recently to study the local environment in Sm and Y doped ceria thin films \cite{Kraynis2019}, Nb-doped TiO$_2$ thin films \cite{Ribeiro2020}, Ir-doped ZnO thin films \cite{Chesnokov2021} and zinc oxide nanopowders \cite{Kuzmin2022}.

RMC/EA simulations were performed using the structural model represented by the 4$a$$\times$4$a$$\times$4$a$ supercell of the zinc blende lattice with the lattice parameter $a$=5.409~\AA\ in agreement with the XRD data (Table \ref{table1}). The supercell for pure ZnS contained 256 zinc and 256 sulphur atoms, whereas 16 zinc atoms were substituted by manganese atoms for Mn-doped ZnS. The Mn atoms were arranged in such a way that the smallest distance between each two of them was more than 10.6~\AA. This arrangement of the Mn atoms ensures that the local environment around each of them can relax independently in RMC simulations. The periodic boundary conditions (PBC) were imposed to avoid surface-related effects, and the largest allowable atom displacements from the position in the ideal zinc blende structure were equal to 0.4~\AA. The number of atomic configurations simultaneously considered in the EA algorithm was 32 \cite{Timoshenko2014}.

At each step of the RMC/EA simulation, the configuration-averaged (CA) EXAFS spectrum was calculated using ab initio real-space multiple-scattering FEFF8.5L code \cite{Ankudinov1998, Rehr2000} over all absorbing atoms of the same type (Zn or Mn) located in the supercell and taking into account the multiple-scattering (MS) effects up to the fourth order. The total number of RMC steps was 5000 to guarantee the convergence of the structural model. The Morlet wavelet transforms (WTs) of the experimental and CA EXAFS spectra $\chi(k)k^2$\ were calculated, and the best agreement between them was used as a criterion for the model structure optimization.  The WT calculations were performed in the k-space range 2.7--15.0~\AA$^{-1}$ at the Zn K-edge and 3.0--12.0~\AA$^{-1}$ at the Mn K-edge. In the R-space the WT calculations were done in the range from 1.0 to 6.0~\AA\ for both absorbing atoms. Note that both Zn and Mn K-edge EXAFS spectra were fitted simultaneously for Mn-doped ZnS nanopowders. 

The atomic coordinates in the final supercell were used to calculate a set of the radial distribution functions (PDFs) $g(r)$ for each sample.  
To improve statistics, ten sets of different (independent) starting structural models were considered for final PDFs. Additionally, the mean-square displacements (MSDs) were evaluated for Zn, S, and Mn atoms, and the mean-square relative displacements (MSRDs) were calculated for Zn--S, Zn--Zn, Mn--S and Mn--Zn atom pairs.

\section{Results and discussion}
\subsection{XRD}

X-ray diffraction patterns of pure and Mn-doped (0.5, 1.0, 1.5, and 2.0 at\%) ZnS nanopowders (Fig.\ \ref{fig1}) correspond to the cubic zinc blende phase with the space group F$\bar{4}$3m (216) and four formula units ($Z$=4) in the unit cell \cite{Jamieson1980}. The zinc atoms occupy the Wyckoff position 4$a$(0,0,0), whereas the sulphur atoms occupy the position 4$c$(0.25,0.25,0.25). 
The Bragg peaks in all patterns are broadened indicating that all samples are nanocrystaline. 
The Rietveld refinement procedure was used to extract the values of the lattice constant $a$ and the average size of crystallites $d$ from X-ray diffraction patterns, which are reported in Table\ \ref{table1}. The lattice constants are in the range of $a$=5.406--5.411~\AA\ with the average value of 5.408(5)~\AA, which is close to that in the bulk zinc blende ZnS ($a$=5.41~\AA\   \cite{Jamieson1980,Skinner1961}) but is smaller than that in zinc blende phase of MnS ($a$=5.59~\AA\ \cite{Mehmed1938}). The average sizes of nanocrystallites are in the range of $d$=6--9~nm.

\subsection{SEM}
\begin{figure}[t]
	\centering
	\includegraphics[width=0.9\linewidth]{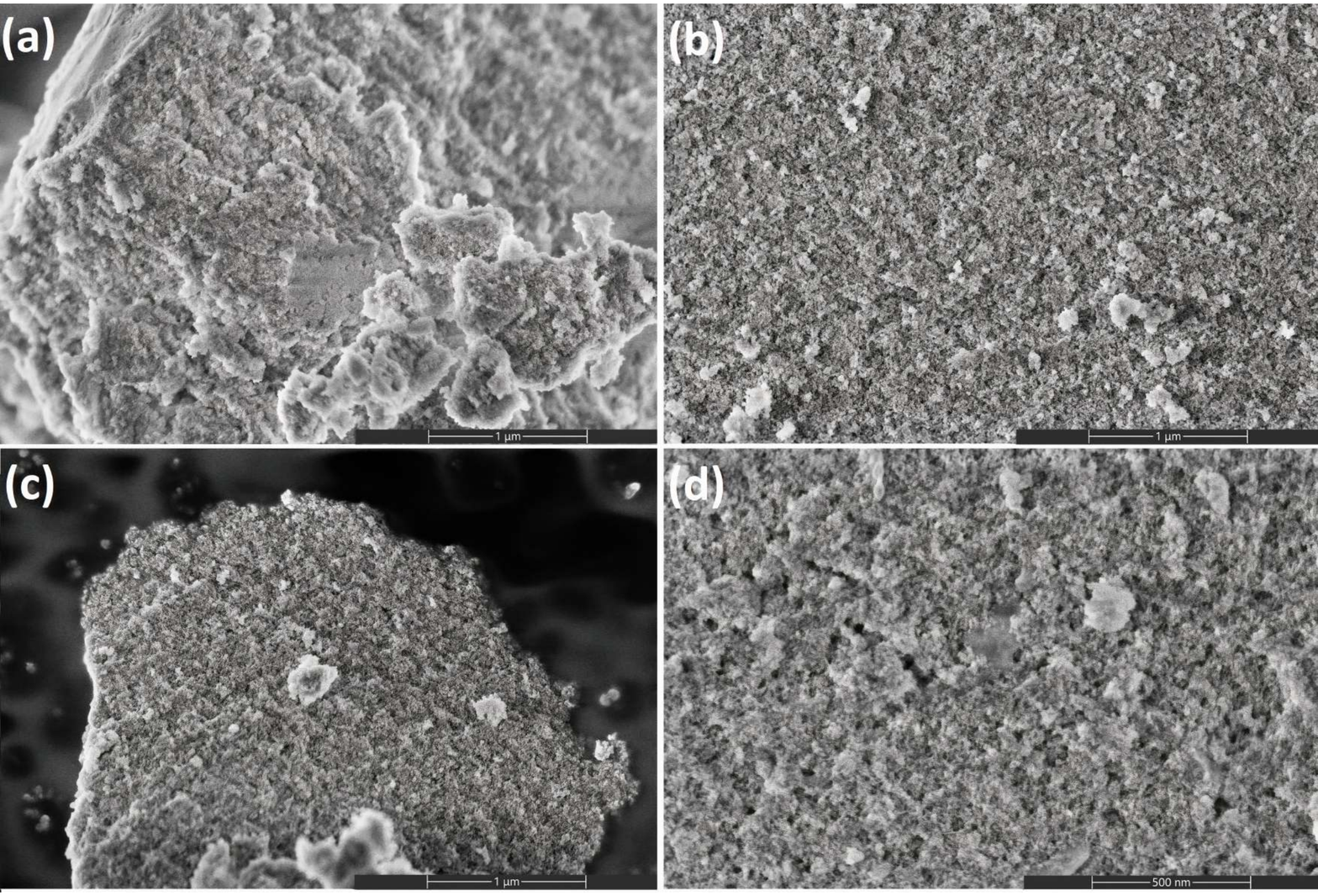}
	\caption{ SEM images of undoped (a) and doped (0.5 at\% (b) and 2.0 at\% (c)) ZnS at 50 kx magnification and  2.0 at\% (d) ZnS:Mn at 100 kx. }
	\label{fig2}
\end{figure}

SEM images provide information on the morphology and microstructure of ZnS:$x$Mn nanoparticles (Fig.\ \ref{fig2}). Although it can be seen that the nanoparticles tend to agglomerate and form large grains with an average size of 10--180~nm, the samples obtained by microwave-assisted solvothermal synthesis showed improved translucency when compared to samples that were previously synthesized without microwave assistance, indicating an overall lower agglomeration. The irregularly shaped agglomerates are formed as a  result of coalescence and can be controlled by optimizing synthesis parameters such as reactor temperature, microwave irradiation time, or adding surfactants \cite{Eggersdorfer2014,Darr2017}. 
Particles with similar morphology have been previously reported in the literature \cite{Mote2013}. The SEM images show no correlation between the concentration of dopant and grain size, nor overall particle morphology. Therefore, these results suggest that the morphology has minimal impact on the photoluminescence data.

The intended Mn-dopant concentration in the synthesized nanoparticles is compared with that obtained by the EDX measurements in Table\ \ref{table1}. The EDX results are in reasonable agreement with the intended dopant concentration except for the 2 at\% sample. However, the photoluminescence measurements show a significant increase in the luminescence intensity for the 2 at\% sample, that together with the intended concentration leads the authors to think that the EDX value for the 2 at\% is undervalued.

\subsection{Photoluminescence}

\begin{figure}[t]
	\centering
	\includegraphics[width=0.9\linewidth]{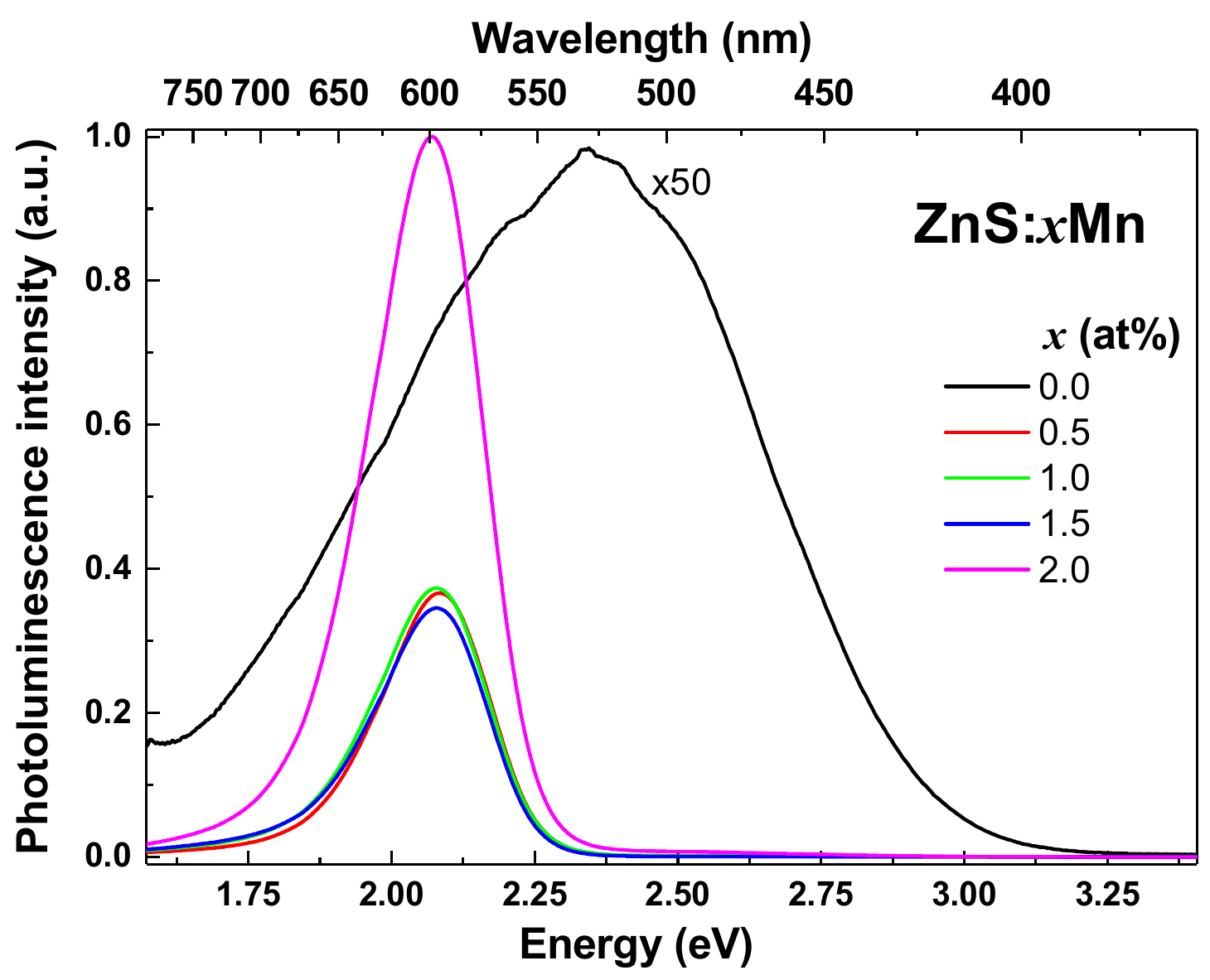}
	\caption{Photoluminescence (PL) emission spectra for ZnS:$x$Mn ZnS powders prepared by 
		microwave-assisted solvothermal synthesis. Note that the PL spectrum for pure ZnS was enlarged by 50 times.  }
	\label{fig3}
\end{figure}

The undoped ZnS sample shows photoluminescence with a broad emission band located at 2.34~eV (530~nm) (Fig.\ \ref{fig3}). The band is attributed to the defects at the surface of the nanoparticles \cite{saleh2019}. Note that the band intensity is low compared to that related to the Mn dopant (see below).  The broadening of the spectrum is explained by two main phenomena -- size distribution and an increase in the surface-to-volume ratio for nanoparticles. 

In all Mn-doped samples, the emission band at 2.07~eV (598~nm) was observed in the PL spectra (Fig.\ \ref{fig3}). 
The Mn$^{2+}$ ligand-field states reside within the optical band gap of ZnS, so the  rapid energy transfer occurs to the Mn$^{2+}$ dopant upon the excitation of the host across the band gap, followed by radiative decay  with the emission of optical photon \cite{Wu2013}.
Therefore, the orange luminescence was assigned to the characteristic forbidden transition between excited ($^4$T$_1$) and ground ($^6$A$_1$) levels within the 3d$^5$ orbital of Mn$^{2+}$. 
The position of the band remains unchanged in all samples indicating that the energies of the Mn internal transitions depend weakly on the composition up to $x$=2 at\%. 
The photoluminescence results suggest that the Mn$^{2+}$ ions occupy Zn$^{2+}$ sites in the host lattice, as was also found in small ZnS:Mn nanoclusters \cite{Sooklal1996} and Zn$_{1-x}$Mn$_x$S nanowires \cite{Brieler2004}.

\subsection{EXAFS}

\begin{figure}[t]
	\centering
	\includegraphics[width=0.95\linewidth]{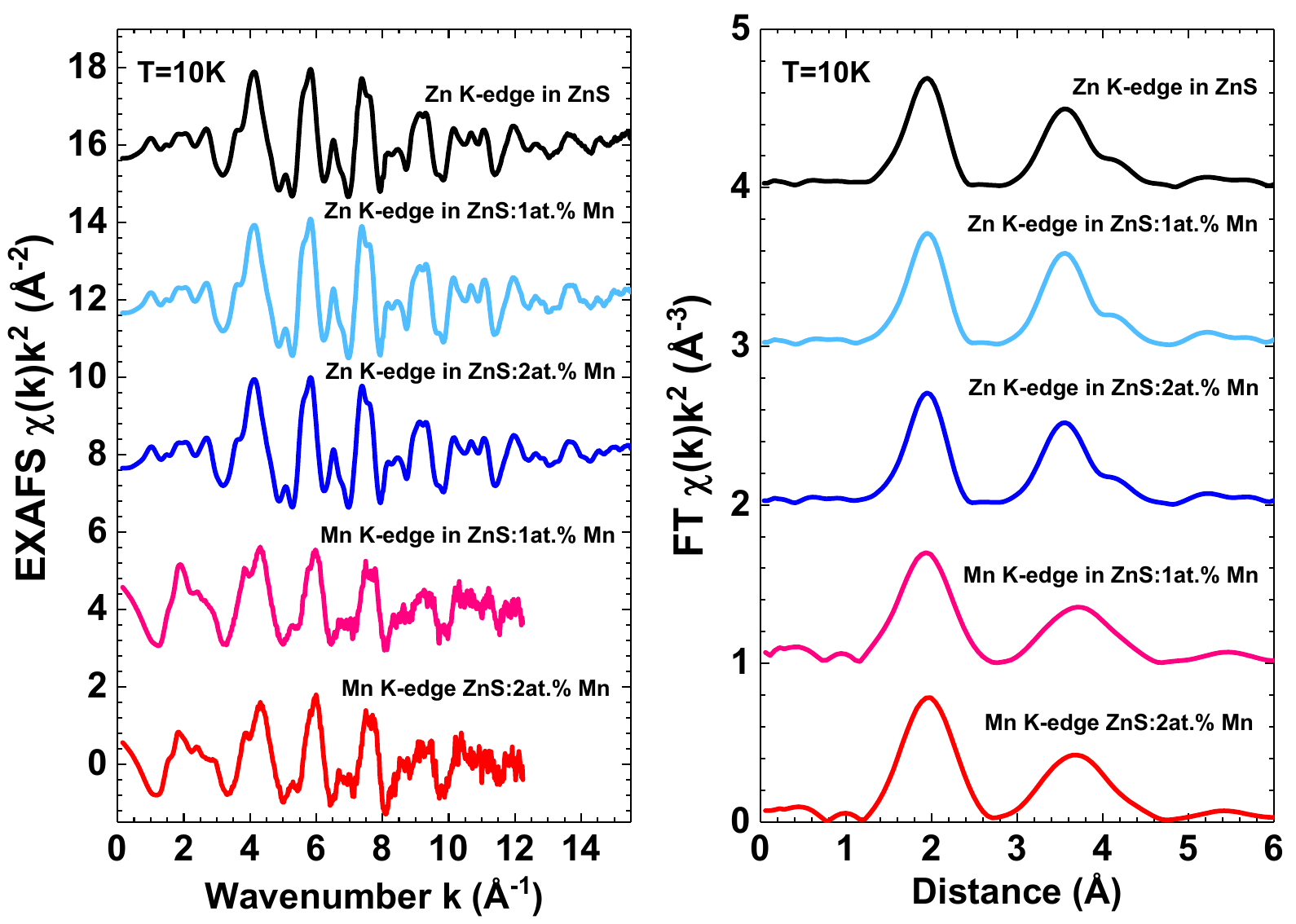}
	\caption{Experimental Zn and Mn K-edge EXAFS spectra $\chi(k)k^2$\ and their Fourier transforms (FTs) for pure and Mn-doped ZnS nanopowder at $T$=10~K. Curves are vertically shifted for clarity. Only moduli of FTs are shown. }
	\label{fig4}
\end{figure}

Experimental Zn and Mn K-edge EXAFS spectra $\chi(k)k^2$\ and their FTs are shown in Fig.\ \ref{fig4} and suggest similarity of the local environment around Zn and Mn atoms in the cubic zinc blende structure. Note that some differences between Zn and Mn K-edge EXAFS spectra at low-$k$ values ($k <3$~\AA$^{-1}$) are due to the difference in the local electronic structure and scattering phase shifts of Zn (3d$^{10}$) and Mn (3d$^2$) absorbing atoms. The FTs are dominated by two peaks located at about 2 and 3.6~\AA. The first peak corresponds to the first coordination shell of metal atoms composed of 4 S atoms, while the second peak has a more complex origin. In pure ZnS, it originates mainly due to the second (12 Zn) and third (12 S) coordination shells of zinc and four MS paths with a total path length less than 5~\AA. The MS paths include two double-scattering contributions (Zn$\rightarrow$S$_1$$\rightarrow$S$_1^\prime$$\rightarrow$Zn*, Zn*$\rightarrow$S$_1$$\rightarrow$Zn$_2$$\rightarrow$Zn*) and two triple-scattering contributions (Zn*$\rightarrow$S$_1$$\rightarrow$Zn*$\rightarrow$S$_1$$\rightarrow$Zn*, Zn*$\rightarrow$S$_1$$\rightarrow$Zn$_2$$\rightarrow$S$_1$$\rightarrow$Zn*): here the asterisk denotes the absorbing atom, the subscript number -- the coordination shell, the accent -- another atom in the same coordination shell. In Mn-doped ZnS, the paths with similar geometry but originating at the Mn absorbing atoms contribute to the Mn K-edge EXAFS.

\begin{figure}[t]
	\centering
	\includegraphics[width=0.6\linewidth]{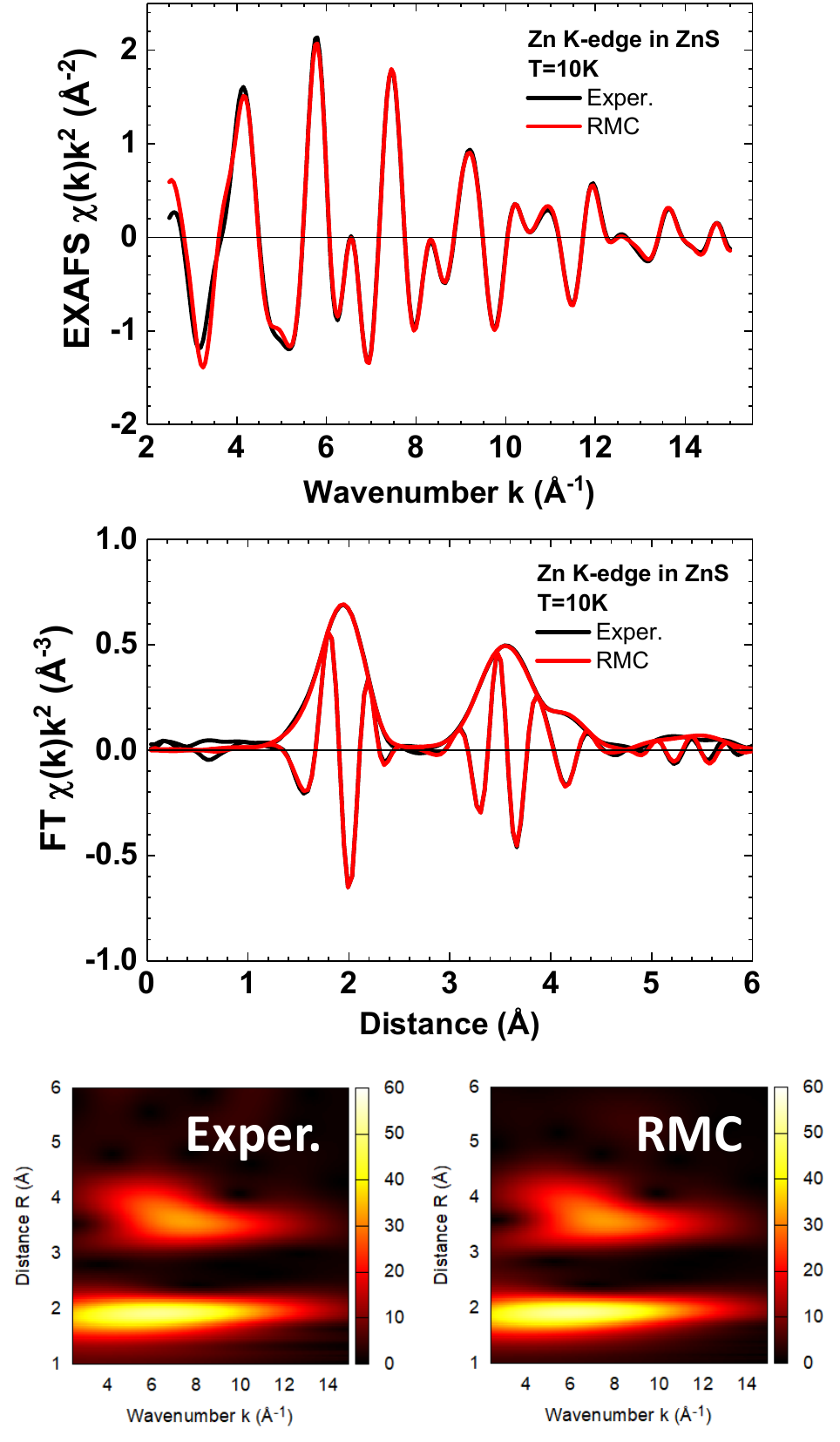}
	\caption{Experimental and RMC-calculated Zn K-edge EXAFS spectra $\chi(k)k^2$\ and their Fourier and Morlet wavelet transforms for ZnS nanopowder at $T$=10~K. Note that the peaks in the Fourier transforms (FTs) are located at distances that are slightly shorter than their crystallographic values because the FTs were not corrected for the phase shift present in the EXAFS equation. }
	\label{fig5}
\end{figure}

\begin{figure}[t]
	\centering
	\includegraphics[width=0.95\linewidth]{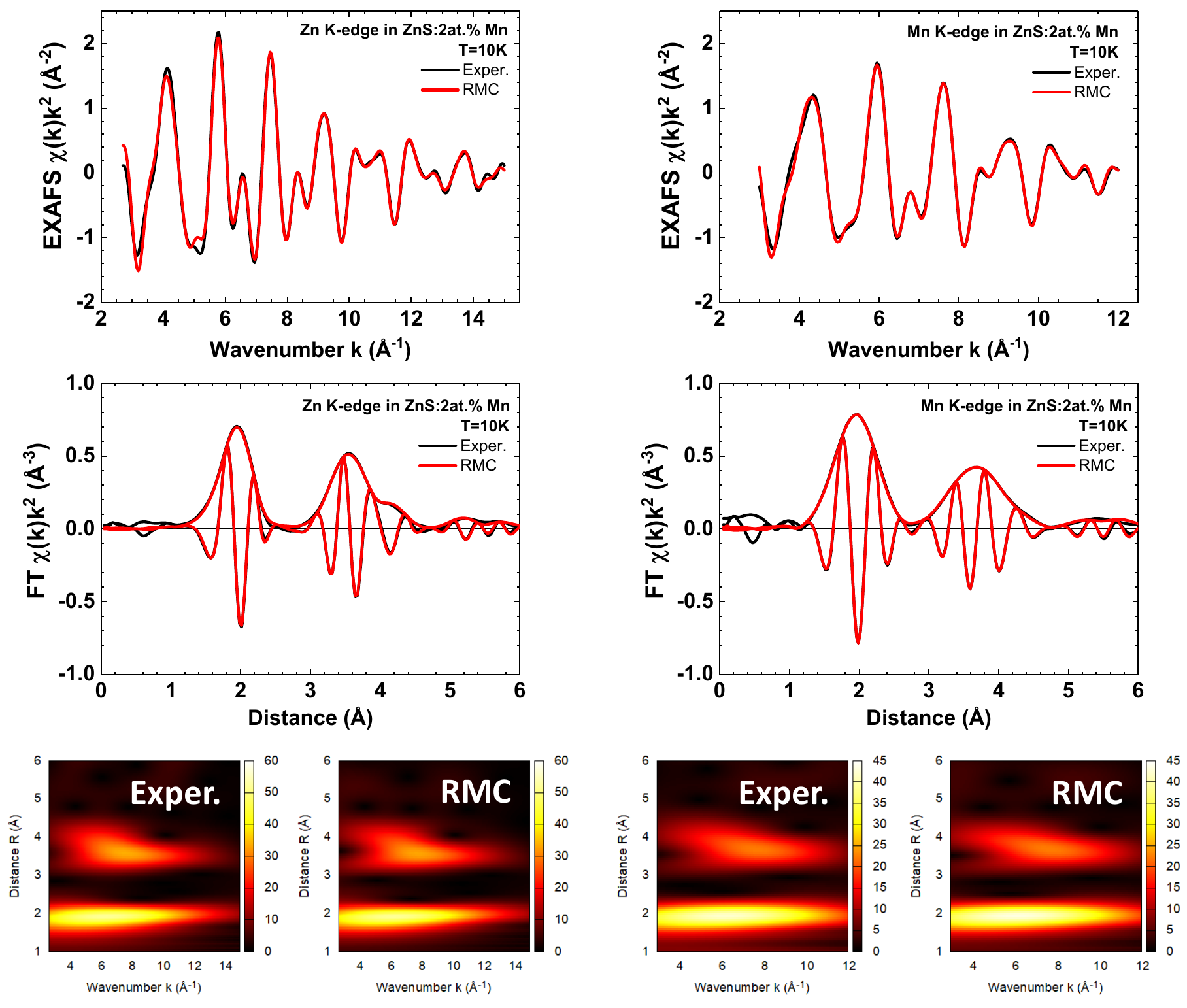}
	\caption{Experimental and RMC-calculated Zn and Mn K-edges EXAFS spectra $\chi(k)k^2$\  and their Fourier and Morlet wavelet transforms for ZnS:2 at\%Mn nanopowder at $T$=10~K. Note that the peaks in the Fourier transforms (FTs) are located at distances that are slightly shorter than their crystallographic values because the FTs were not corrected for the phase shift present in the EXAFS equation.}
	\label{fig6}
\end{figure}

\begin{figure}[t]
	\centering
	\includegraphics[width=0.6\linewidth]{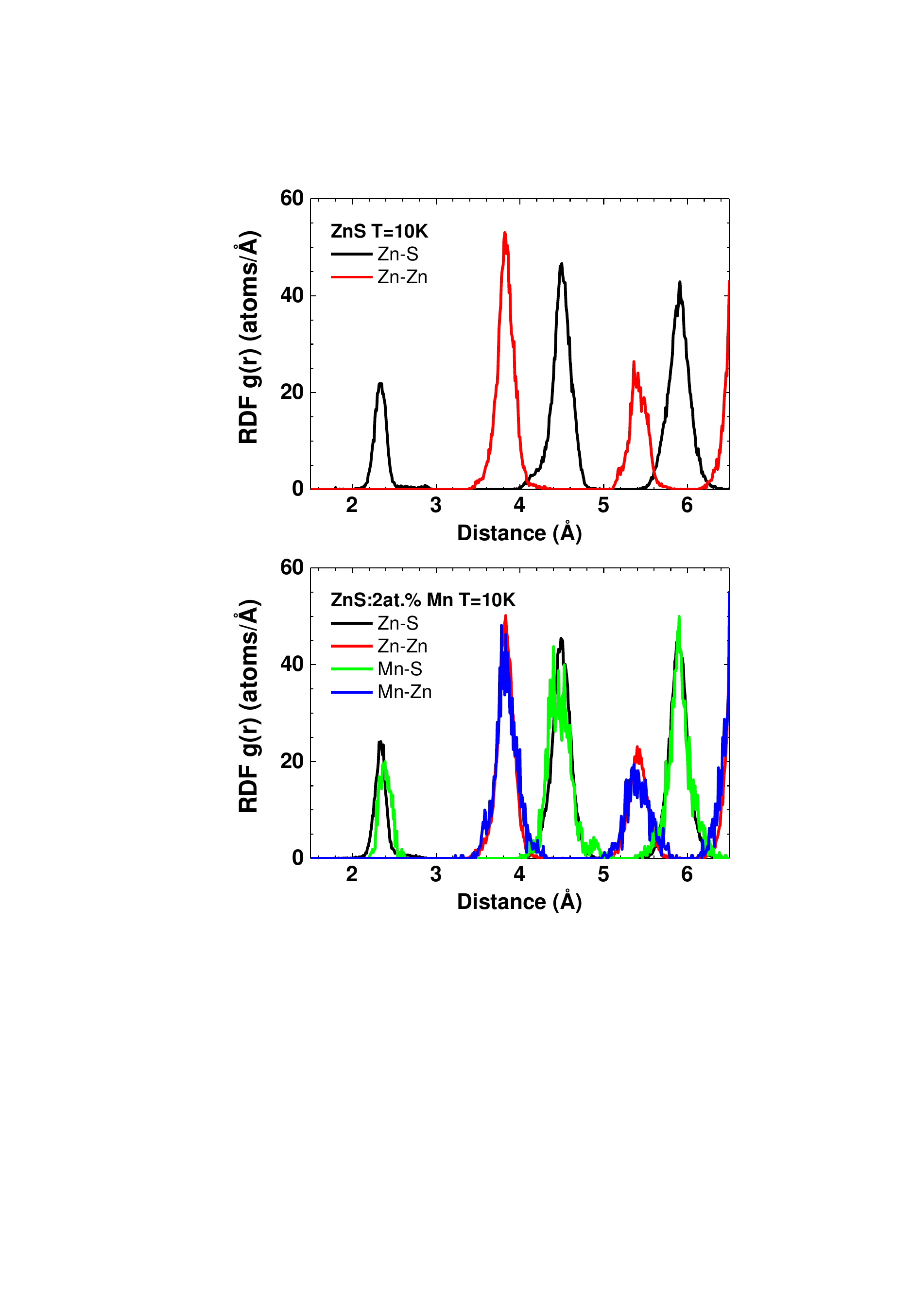}
	\caption{Radial distribution functions (RDFs) $g(r)$ for Zn--S, Zn--Zn, Mn--S and Mn--Zn atom pairs in ZnS and ZnS:2 at\%Mn at $T$=10~K.  }
	\label{fig7}
\end{figure}

The results of the RMC/EA simulations for ZnS and ZnS:2 at\% Mn reported in Figs.\ \ref{fig5} and \ref{fig6} show good agreement between model and experiment in both $k$ and $R$ spaces. A similar agreement was found in the case of ZnS:1 at\% Mn sample (not shown here). Thus, both nanopowders have a cubic zinc blende structure, and manganese atoms substitute zinc with tetrahedral coordination by 4 sulphur atoms. The analysis of the partial RDFs $g(R)$ (Fig.\ \ref{fig7}) provides more detailed and qualitative information. The interatomic distance $R$(Mn--S) is equal to 2.40(2)~\AA\ being larger than $R$(Zn--S)=2.33(2)~\AA. 
These values agree well with that found previously by EXAFS in Zn$_{1-x}$Mn$_x$S nanoparticles \cite{Soo1994,Dinsmore2000,Cao2010} and nanowires \cite{Brieler2004}.
Note also that Mn--S distance is close to that in bulk zinc blende MnS  ($R$(Mn--S)=2.42~\AA\ \cite{Mehmed1938}).
The difference between the two distances agrees well with the difference in the ionic radii for two ions in a tetrahedral environment: the ionic radius of Zn$^{2+}$(IV) is 0.60~\AA, while that of Mn$^{2+}$(IV) is 0.66~\AA\ \cite{Shannon1976}.

The MSRDs calculated for the  first and second coordination shells of metal atoms from the RDFs in Fig.\ \ref{fig7} are equal to $\sigma^2$(Zn--S) = 0.005(3)~\AA$^2$, $\sigma^2$(Zn--Zn) = 0.010(3)~\AA$^2$, $\sigma^2$(Mn--S) = 0.006(3)~\AA$^2$ and $\sigma^2$(Mn--Zn) = 0.018(3)~\AA$^2$. They have close values for the Zn and Mn environment, which additionally supports the substitution of Zn by Mn. 

The MSD values calculated from the coordinates of atoms in the RMC supercell relative to their ideal positions in the zinc blende structure can be used to estimate the degree of static disorder present in the samples due to their nanocrystalline nature. In small nanocrystals ($<10$~nm), an increase of MSDs is expected due to the relaxation of the outermost layers located close to the nanocrystal surface.
Additional effects can occur due to a deviation from stoichiometry and the local structure relaxation induced by substitution. The MSD values for Zn, S, and Mn atoms and the MSRDs for Zn--Zn atom pair with the interatomic distance $R$$\approx$5.9~\AA\ are reported in Table\ \ref{table2}. 
Note that there is a relationship between MSRD and MSD for a pair of atoms $i$ and $j$: MSRD$_{ij}$ = MSD$_i$ + MSD$_j$ - 2$\varphi$(MSD$_i$)$^{1/2}$(MSD$_j$)$^{1/2}$, where $\varphi$\ is a dimensionless correlation parameter \cite{Jeong2003}. For large interatomic distances ($\geq$5--6~\AA\ \cite{Jeong2003, Jonane2018, Bocharov2021}), the correlation in the motion of atoms disappears, i.e. $\varphi \rightarrow 0$, and MSRD is equal to the sum of two MSDs. As one can see in Table\ \ref{table2}, 2$\times$MSD(Zn) are large than  MSRD(Zn--Zn) by about two times, suggesting that there is an additional contribution to each MSD factor that is absent in the MSRD. This additional contribution comes from the procedure for estimating the MSDs with respect to the ideal position of atoms in the zinc blende structure. Indeed, some corrections should be made for atoms located close to the surface of nanocrystals. Thus, large MSD values reflect the nanocrystallinity of our samples.

\begin{table}[t]\footnotesize
	\caption{The mean-square displacements (MSDs) for Zn, S, and Mn atoms and the mean-square relative displacement (MSRD) for Zn--Zn atom pair with the interatomic distance $R \approx 5.9$~\AA\ in ZnS:$x$Mn nanopowders obtained from the RMC simulations. Values of MSD and MSRD are in \AA$^2$.}  \label{table2}
	\renewcommand{\arraystretch}{1.2} 
	\begin{tabular}{@{}lcccc}
		\hline
		$x$ (at.\%)	& MSD(Zn) 	& MSD(Mn)	 & MSD(S)  & MSRD(Zn--Zn) \\
		\hline
		0	& 0.015	&--		&0.025	&0.011	\\	
		1.0	& 0.014	&0.030	&0.020	&0.010	\\
		2.0	& 0.016	&0.033	&0.021	&0.011	\\
		\hline
	\end{tabular}\\[2pt]
\end{table}

\section{Conclusions}
Undoped and Mn-doped (0.5, 1.0, 1.5, 2.0 at\%)  zinc sulfide nanoparticles were synthesized by the microwave-assisted solvothermal (MWST) method.  They were characterized by  X-ray diffraction, photoluminescence spectroscopy, and scanning electron microscopy with energy-dispersive X-ray spectroscopy. The local atomic structure of nanoparticles and its relaxation due to doping was probed by X-ray absorption spectroscopy at the Zn and Mn  K-edges. 

Rietveld refinement of XRD patterns suggested that all samples are nanocrystalline with the cubic zinc blende structure. The average lattice parameter is $a$=5.408(5)~\AA, and the average size of crystallites  is in the range of 6--9~nm. Nevertheless, the SEM study evidenced the tendency of nanoparticles to agglomerate and form large grains with an average size of up to 180~nm. 

Analysis of the Zn and Mn K-edge EXAFS spectra using the reverse Monte Carlo method confirmed that manganese ions substitute zinc ions. The difference in ion sizes is responsible for the larger interatomic distances Mn--S (2.40(2)~\AA) compared to Zn--S (2.33(2)~\AA). The local structure relaxation due to the small size of crystallites is responsible for the large values of the MSD factors for Zn, S and Mn atoms, calculated relative to the ideal position of atoms in the zinc blende structure.

\section*{CRediT authorship contribution statement}
\textbf{Alexei Kuzmin}: Investigation, Conceptualization, Validation, Formal analysis, Writing – original draft, Writing – review \& editing, Supervision, Funding acquisition. 
\textbf{Milena Dile}: Investigation, Conceptualization, Validation, Writing – original draft, Writing – review \& editing. 
\textbf{Katrina Laganovska}: Investigation, Conceptualization, Validation, Writing – original draft, Writing – review \& editing. 
\textbf{Aleksejs Zolotarjovs}: Investigation, Conceptualization, Resources, Writing – original
draft, Writing – review \& editing, Supervision.

\section*{Declaration of Competing Interest}
The authors declare that they have no known competing financial interests or personal relationships that could have appeared to influence the work reported in this paper.

\section*{Data availability}
Data associated with this article are available upon reasonable request to the authors.

\section*{Acknowledgements}
The financial support of the European Regional Development Fund (ERDF) Project No. 1.1.1.1/20/A/060 is greatly acknowledged. 
We acknowledge DESY (Hamburg, Germany), a member of the Helmholtz Association HGF, for the provision of experimental facilities. Parts of this research were carried out at PETRA III and we would like to thank Dr. Edmund Welter for assistance in using P65 beamline. Beamtime was allocated for proposal I-20210366 EC.
Institute of Solid State Physics, University of Latvia as the Center of Excellence has received funding from the European Union's Horizon 2020 Framework Programme H2020-WIDESPREAD-01-2016-2017-TeamingPhase2 under grant agreement No. 739508, project CAMART2.






\end{document}